\documentclass[onecolumn,preprint]{pasj00}%

\begin{document}

\SetVolumeData{}{??}{?}
\SetRunningHead{Yamamoto, Nakamichi, Kamino, Bassett and Nishioka}{}

\title{A Measurement of the Quadrupole Power Spectrum \\in the Clustering 
of the 2dF QSO Survey}
\author{Kazuhiro \textsc{Yamamoto}, Masashi \textsc{Nakamichi}, 
Akinari \textsc{Kamino}}%
\affil{Graduate School of Science, Hiroshima University, 
Higashi-Hiroshima, 735-8526, Japan}

\author{Bruce A. \textsc{Bassett}}
\affil{Institute of Cosmology and Gravitation, University of Portsmouth, England and\\ 
SAAO, Observatory, Cape Town, South Africa}

\author{Hiroaki \textsc{Nishioka}}
\affil{Institute of Astronomy and Astrophysics, 
Academia Sinica, Taipei 106, Taiwan, R.O.C.}


\maketitle

\begin{abstract}
We report a measurement of the quadrupole power spectrum in 
the two degree field (2dF) QSO redshift (2QZ) survey. The analysis uses 
an algorithm parallel to that for the estimation of the standard monopole
power spectrum without first requiring computation of the correlation 
function or the anisotropic power spectrum. 
The error on the quadrupole spectrum is rather large but the best fit 
value of the bias parameter from the quadrupole spectrum is 
consistent with that from previous investigations of the 2dF data.
\end{abstract}

\section{Introduction}
\def\M{{M}}
\def\w{{\psi}}
\def\dls{{D_{\rm LS}}}
\def\dos{{D_{\rm OS}}}
\def\bftheta{{\Theta}}
\def\calD{{\cal D}}
\def\bfk{{\bf k}}
\def\bfs{{\bf s}}

Since large cosmological  surveys are carried out in redshift space,
the peculiar velocity of sources distorts the apparent spatial distribution of 
cosmological objects. These redshift-space 
distortions are one of the important effects in large redshift 
surveys (e.g., Suto \etal~2000a, Magira, Jing, Suto 2000). 
Redshift distortions can be classified into the linear distortion 
arising in the linear theory of density perturbation and 
the {\em Finger of God} effect arising  
in the nonlinear regime. It has been also pointed out that the 
the geometrical effect from the expansion of the universe
causes another apparent distortion in the distribution of 
cosmological objects, which is referred to as the 
cosmological redshift-space distortion or the geometric distortion
(\cite{AP}, \cite{BPH}, \cite{MS}).
Thus the detection of the redshift-space distortion provides
us with unique information about the peculiar velocity and
expansion history of the universe. 

In the Kilo-Aperture Optical Spectrograph (KAOS/WFMOS, see http://www.noao.edu/kaos/),
one of the future large survey projects, it will be possible to confront in great detail 
the theoretical redshift-space 
distortions including the geometric distortion with observation (e.g., Yamamoto, Nishioka, Bassett 2005).
Such surveys will provide significant dark energy constraints (Matsubara, Szalay 2003, 
Seo, Eisenstein 2003, Linder 2003, 
Blake, Glazebrook 2003, Yamamoto 2003;2004, Amendola \etal 2005) using the
baryon acoustic signatures, which have recently been detected clearly in the SDSS and 2df data sets.
(Eisenstein \etal~2005, Cole \etal~2005, Yahata \etal~2005). 

In general it is known that the redshift-space power spectrum 
on a constant hypersurface of the redshift $z$ can be expanded 
as (\cite{TH}) 
\begin{eqnarray}
  P({\bf k},z)=P(k,\mu,z)=\sum_{l=0,2,4\cdots} 
  P_{l}(k,z){\cal L}_{l}(\mu)(2l+1),
\label{pkzexpand}
\end{eqnarray}
where ${\cal L}_l(\mu)$ is the Legendre polynomials, 
$\mu(=\cos\theta)$ is the directional cosine between the line of sight
direction and $\bfk$ (See Figure~1). \footnote{Note that our definition of 
the multipole spectrum $P_{l}$ is different from the conventional  definition by the factor $2l+1$.}
The monopole $P_0(k,z)$ represents the angular averaged power 
spectrum, what we mean by the power spectrum usually, and 
the quadrupole $P_2(k,z)$ represents the leading 
anisotropies in the power spectrum due to the redshift-space 
distortions. 
The odd moments vanish by symmetry.
The quadrupole spectrum provides us with new
information over what is available in the monopole power spectrum. 
In practice, the quadrupole spectrum reflects the peculiar and
random velocity of samples (Kaiser 1987, Cole, Fisher, Weinberg 1994, Hamilton 1996). 
It has been shown that the quadrupole spectrum will be 
especially useful in breaking the degeneracy 
between the bias and the dark energy parameters when measurements of 
the power spectrum are good enough in future surveys (Yamamoto \etal~2005).
Thus accurate measurement of the quadrupole spectrum can be 
important in characterizing the redshift-space distortions 
quantitatively. 

Pioneering work on the measurement of the quadrupole power 
spectrum was carried out by Cole, Fisher and Weinberg (1994) and
Hamilton (1996, 1997) using galaxy redshift survey catalogs. 
Cole et~al. (1994) presented a systematic method
to estimate the quadrupole power spectrum through the 
anisotropic power spectrum. Using the method,
Hatton and Cole (1999) estimated the $\beta$-factor from
the quadrupole power spectrum in the two degree field (2dF) 
galaxy survey. 
In the work by Hamilton (1996,1997) the quadrupole power spectrum 
was obtained by the transformation of the correlation function. 
In the present work, however, we consider a different method to 
estimate the higher multipole moments of the power spectrum. 
Our method is parallel to the one widely used to measure the monopole 
power spectrum (Feldman, Kaiser, Peacock 1994, Yamamoto 2003) 
and allows us to obtain the multipoles of the redshift-space 
power spectrum without evaluating the correlation function or 
the anisotropic power spectrum. 
The difference between our method and that developed by 
Cole \etal (1994) is discussed in the next section.

This paper is organized as follows: In section 2, we explain 
the algorithm for estimating the multipole moments of the 
power spectrum. In section 3, we apply the method to the 2dF 
QSO sample and obtain the quadrupole power spectrum. Using 
the result we briefly discuss the constraint on the bias parameter. 
The last section is devoted to summary and discussions. 
Throughout this paper we use units in 
which the speed of light is unity, $c=1$.

\section{Method}
Here we explain the optimal weighting scheme for estimating the multipole
moments of the anisotropic (i.e. redshift-space) power spectrum. 
Our method developed here is unique and different
from those developed previously. 
Our method does not require computation of the correlation function; c.f. Hamilton (1996). 
A similar method was developed by Cole et~al. (1994), which also does not 
require the computation of the correlation function. However it 
requires estimating the anisotropic power spectrum.  The difference 
between our method and their method will be discussed in detail below,
however, the essence is as follows. 

In the Cole et~al. method, 
it is necessary to divide the sample into subsamples and introduce
one line of sight direction for each subsample in order to 
explicitly define $\mu$.
Thus in the Cole et~al method, the line of sight directions for all objects are regarded 
as the same in each divided subsample. In our method, however, we 
need neither divide the sample nor explicitly define $\mu$, instead, 
we introduce $\mu$ for each pair of objects of the sample  (see Figure 1). 
Furthermore, the multipole moments can be computed fully in parallel with the 
conventional monopole estimation (i.e. the 'usual power spectrum').

Throughout this paper, we use $\bfs$ to denote the three 
dimensional coordinates in redshift space. 
We also use $n_g(\bfs)$ and $\bar n(\bfs)$ to denote real catalog and 
the expected mean number density. A fluctuation field is then defined via 
\begin{eqnarray}
  F(\bfs)=n_g(\bfs)-\alpha n_s(\bfs),
\end{eqnarray}
where $n_g(\bfs)=\sum_i \delta(\bfs- \bfs_i)$ with $\bfs_i$ being
the location of the $i$th object of real catalog; 
similarly $n_s(\bfs)=\sum_j \delta(\bfs- \bfs_j)$ is the
density of a synthetic catalog. The synthetic catalog is 
a set of random points without correlation, which can be
constructed through random process by mimicking the 
selection function of the real catalog. 
Here we assume that the synthetic catalog has a mean number 
density $1/\alpha$ times that of the real catalog. 
The synthetic catalog provides an estimate for the mean 
number density in the absence of clustering, then it is 
useful to provide an estimator for the fluctuation field 
$F(\bfs)$. 
{}From the definition of the random field $n_s(\bfs)$,
we assume (\cite{FKP}),
\begin{eqnarray}
  &&\langle n_g(\bfs_1)n_g(\bfs_2)\rangle=
  \bar n(\bfs_1)\bar n(\bfs_2)(1+\xi(\bfs_1,\bfs_2))
  +\bar n(\bfs_1)\delta(\bfs_1-\bfs_2),
\label{nnA}
\\ 
 &&\langle n_s(\bfs_1)n_s(\bfs_2)\rangle=
  \alpha^{-2}\bar n(\bfs_1)\bar n(\bfs_2)
  +\alpha^{-1}\bar n(\bfs_1)\delta(\bfs_1-\bfs_2),
\label{nnB}
\\ 
 &&\langle n_g(\bfs_1)n_s(\bfs_2)\rangle=
  \alpha^{-1}\bar n(\bfs_1)\bar n(\bfs_2),
\label{nnC}
\end{eqnarray}
where $\xi(\bfs_1,\bfs_2)$ denotes the two-point correlation function.
%
%
Using these relations, we have
\begin{eqnarray}
  \langle F(\bfs_1)F(\bfs_2)\rangle
  =\bar n(\bfs_1)\bar n(\bfs_2)\xi(\bfs_1,\bfs_2)
  +(1+\alpha)\bar n(\bfs_1)\delta(\bfs_1-\bfs_2) .
\end{eqnarray}

The estimator for the multipole moments of the power spectrum 
may be defined as follows (see Yamamoto, Nishioka, Taruya 2000),
\begin{eqnarray}
  R_{l}(\bfk)={\int d\bfs_1 \int d\bfs_2 \w(\bfs_1,\bfk)\w(\bfs_2,\bfk)
  F(\bfs_1)F(\bfs_2)e^{i\bfk\cdot (\bfs_1-\bfs_2)}
  {\cal L}_{l}(\hat \bfs_h\cdot \hat \bfk)\over
  \int d\bfs \bar n^2(\bfs) \w(\bfs,\bfk)^2},
\label{defff}
\end{eqnarray}
where $\bfs_h=(\bfs_1+\bfs_2)/2$ and the `hat' means the unit vector, 
and $\w(\bfs,\bfk)$ is a weight factor. As mentioned before, 
${\cal L}_l(\mu)$ is the Legendre polynomials. (See also Figure 1 for the 
definition of variables). We can choose the weight factor so as 
to minimize the variance of the spectrum in order to optimize 
the performance of the estimator according to the local density 
field (see below).

Here we adopt the approximate formula
\begin{eqnarray}
  \xi(\bfs_1,\bfs_2)=\int {d\bfk \over (2\pi)^3}P(\bfk,|\bfs_h|)
  e^{-i\bfk\cdot(\bfs_1-\bfs_2)},
\label{defxixi}
\end{eqnarray} 
where $P(\bfk,|\bfs_h|)$ is 
the power spectrum defined on a constant-time hypersurface.
This expression is not strictly correct when  
the redshift $\bfs_1$ and $\bfs_2$ are very different. 
Thus (\ref{defxixi}) is justified within the distant observer approximation,
$|\bfs_1-\bfs_2|\ll |\bfs_1|, |\bfs_2|$, i.e., 
in case of short separation between $\bfs_1$ and $\bfs_2$. 

Using the above relations and under the distant observer approximation, 
the ensemble average of $R_{l}(\bfk)$ is 
\begin{eqnarray}
  \langle R_{l}(\bfk)\rangle=
  {\int d\bfs \bar n(\bfs)^2\w(\bfs,\bfk)^2 P(\bfk,|\bfs|)
  {\cal L}_{l}(\hat \bfs\cdot \hat \bfk)
\over
  \int d\bfs \bar n^2(\bfs) \w(\bfs,\bfk)^2}
  +S_{l}(\bfk),
\end{eqnarray}
where 
\begin{eqnarray}
  S_{l}(\bfk)  ={ (1+\alpha) \int d\bfs {\bar n}(\bfs) \w(\bfs,\bfk)^2
  {\cal L}_{l}(\hat \bfs\cdot \hat \bfk)
\over
  \int d\bfs {\bar n}^2(\bfs) \w(\bfs,\bfk)^2}.
\end{eqnarray}
The estimator of the anisotropic power spectrum is obtained 
by subtracting the shotnoise $S_{l}(\bfk)$, yielding
\begin{eqnarray}
  {\cal P}_{l}(\bfk)=R_{l}(\bfk) -S_{l}(\bfk).
\end{eqnarray}
Next we focus on the covariance of ${\cal P}_{l}(\bfk)$, which is given by
\begin{eqnarray}
  \langle\Delta{\cal P}_{l}(\bfk)\Delta{\cal P}_{l}(\bfk')\rangle
  &=&\langle[{\cal P}_{l}(\bfk)-\langle{\cal P}_{l}(\bfk)\rangle]
    [{\cal P}_{l}(\bfk')-\langle{\cal P}_{l}(\bfk')\rangle]\rangle
\nonumber\\
  &=&\langle{\cal P}_{l}(\bfk){\cal P}_{l}(\bfk')\rangle
    -\langle{\cal P}_{l}(\bfk)\rangle\langle{\cal P}_{l}(\bfk')\rangle
\nonumber\\
  &=&\langle{R}_{l}(\bfk){R}_{l}(\bfk')\rangle
    -\langle{R}_{l}(\bfk)\rangle\langle{R}_{l}(\bfk')\rangle.
\end{eqnarray}
Assuming Gaussian statistics some calculation gives:
\begin{eqnarray}
   \langle F(\bfs_1)F(\bfs_2)F(\bfs_3)F(\bfs_4)\rangle &=&
   \langle F(\bfs_1)F(\bfs_2)\rangle
   \langle F(\bfs_3)F(\bfs_4)\rangle
\nonumber
\\
   &+&\langle F(\bfs_1)F(\bfs_3)\rangle
   \langle F(\bfs_2)F(\bfs_4)\rangle
\nonumber
\\
  &+&\langle F(\bfs_1)F(\bfs_4)\rangle
   \langle F(\bfs_3)F(\bfs_2)\rangle,
\end{eqnarray}
we have
\begin{eqnarray}
  \langle \Delta {\cal P}_{l}(\bfk) \Delta {\cal P}_{l}(\bfk') \rangle&=&
  [\int d\bfs \bar n(\bfs)^2 \w(\bfs,\bfk)^2]^{-1}
  [\int d\bfs'\bar n(\bfs')^2\w(\bfs',\bfk')^2]^{-1}
\nonumber
\\
  && \times 2 \prod_{i=1}^2 [\int d\bfs_i \w(\bfs_i,\bfk)]
              \prod_{j=3}^4 [\int d\bfs_j \w(\bfs_j,\bfk')]
   \langle F(\bfs_1)F(\bfs_3)\rangle
   \langle F(\bfs_2)F(\bfs_4)\rangle
\nonumber
\\
   && \times e^{i\bfk\cdot(\bfs_1-\bfs_2)}e^{-i\bfk'\cdot(\bfs_3-\bfs_4)}
   {\cal L}_{l}(\hat \bfs_h\cdot \hat \bfk)
   {\cal L}_{l}(\hat \bfs_h'\cdot \hat \bfk'),
\label{defPPB}
\end{eqnarray}
where $\bfs_h=(\bfs_1+\bfs_2)/2$ and $\bfs_h'=(\bfs_3+\bfs_4)/2$.
Repeatedly using the distant observer approximation, 
\begin{eqnarray}
  &&\int d\bfs_1\int d\bfs_3 \w(\bfs_1,\bfk) \w(\bfs_3,\bfk')
  \langle F(\bfs_1)F(\bfs_3)\rangle
  e^{i\bfk\cdot\bfs_1-i\bfk'\cdot\bfs_3}
\nonumber
\\
  &&\hspace{2cm}\simeq \int d \bfs \bar n(\bfs)^2 
  \w(\bfs,\bfk)\w(\bfs,\bfk')
  e^{i\bfs\cdot(\bfk-\bfk')} 
  \Bigl[
  P\Bigl({\bfk+\bfk'\over 2},|\bfs|\Bigr)
  +{\alpha+1 \over \bar n(\bfs)}\Bigr],
\end{eqnarray}
we finally have 
\begin{eqnarray}
  &&\langle \Delta {\cal P}_{l}(\bfk) \Delta {\cal P}_{l}(\bfk') \rangle
  \simeq
  \overline {\Delta P^2_{l}}(\bfk) \delta^{(3)}(\bfk-\bfk'),
\label{defPC}
\end{eqnarray}
where
\begin{eqnarray}
  &&  \overline {\Delta P^2_{l}}(\bfk)=
  2(2\pi)^3
  {\int d\bfs \bar n(\bfs)^4 \w(\bfs,\bfk)^4\left[
  P(\bfk,|\bfs|)+{(1+\alpha)/ \bar n(\bfs)}\right]^2
  [{\cal L}_{l}(\hat \bfk\cdot \hat \bfs)]^2
  \over 
  [\int d\bfs' \bar n(\bfs')^2 \w(\bfs',\bfk)^2]^2}.
\end{eqnarray}

We define the estimator of the multipole moments of the 
anisotropic power spectrum by
\begin{eqnarray}
  &&{\cal P}_{l}(k)={1\over \Delta V_{k} }
  \int_{\Delta V_{k}} d\bfk {\cal P}_{l}(\bfk),
\label{defcalPzero}
\end{eqnarray}
where $\Delta V_{k}$ denotes the volume of the shell
in the Fourier space. 
The ensemble average of ${\cal P}_{l}(k)$ is given by
\begin{eqnarray}
  \langle {\cal P}_{l}(k)\rangle=
  {1\over \Delta V_{k}}\int_{\Delta V_{k}} d\bfk
  {\int d\bfs \bar n(\bfs)^2\w(\bfs,\bfk)^2 P(\bfk,|\bfs|)
  {\cal L}_{l}(\hat \bfk\cdot \hat \bfs)
\over
  \int d\bfs' \bar n^2(\bfs') \w(\bfs',\bfk)^2}.
\label{expll}
\end{eqnarray}
Note that ${\cal P}_{l}(k)$ corresponds to the multipole
coefficient $P_{l}(k,z)$ in equation (\ref{pkzexpand}). 
${\cal P}_{l}(k)$ is generalized so as to incorporate 
the redshift evolution. 
In the limit that the sources are confined to a very narrow range of
redshifts $z$, $\langle {\cal P}_{l}(k)\rangle$ reduces to 
$P_{l}(k,z)$.

The variance of ${\cal P}_{l}(k)$ is 
obtained by evaluating
\begin{eqnarray}
  \langle \Delta {\cal P}_{l}(k)^2 \rangle &\equiv&
  \langle [{\cal P}_{l}(k)-  \langle{\cal P}_{l}(k)
  \rangle]^2 \rangle
\nonumber
\\
  &=& {1\over \Delta V_{k}^2}  
  \int_{\Delta V_k} d\bfk \int_{\Delta V_k} d\bfk'
  \langle \Delta {\cal P}_{l}(\bfk) \Delta {\cal P}_{l}(\bfk') \rangle,
\label{defPPE}
\end{eqnarray}
which reduces to
\begin{eqnarray}
   \langle \Delta {\cal P}_{l}(k)^2 \rangle &=&
  2{(2\pi)^3\over \Delta V_{k}}
  {\cal Q}^2_{l}(\bfs,k),
\label{defPPF}
\end{eqnarray}
where we have defined
\begin{eqnarray}
  &&{\cal Q}^2_{l}(\bfs,k) = {1\over \Delta V_{k}}
  \int_{\Delta V_k} d\bfk   {
  \int d\bfs \bar n(\bfs)^4 
 \w(\bfs,\bfk)^4 \bigl[P\bigl({\bfk},|\bfs|\bigr)
  +{1/\bar n(\bfs)}\bigr]^2
  [{\cal L}_{l}(\hat \bfk\cdot \hat \bfs)]^2
  \over
  [\int d\bfs' \bar n(\bfs')^2 \w(\bfs',\bfk)^2]^2}.
\label{defPPG}
\end{eqnarray}
Here we have assumed  $\alpha\ll1$. 

Now we find that the following weight factor gives the stationary 
solution 
\footnote{
Compare with the result in (Yamamoto 2003), 
in which the weighting factor was determined 
under the constraint that it has no angular dependence 
$\psi(\bfs,\bfk)=\psi(\bfs,k)$.}
\begin{eqnarray}
  \w(\bfs,\bfk)={1\over 1+\bar n(\bfs) P(\bfk,|\bfs|)}
\label{defoptw}
\end{eqnarray}
for the variation 
\begin{eqnarray}
{\delta \langle \Delta {\cal P}_{l}(k)^2 \rangle
\over \delta \w(\bfs,\bfk)}=0.
\end{eqnarray}
In this case find the minimum error
\begin{eqnarray}
   && \langle \Delta {\cal P}_{l}(k)^2 \rangle=
  2{(2\pi)^3\over \Delta V_{k}}
  {1\over 2}
  \int_{-1}^1 d\mu  
  {[{\cal L}_{l}(\mu)]^2\over
  \int d\bfs' \bar n(\bfs')^2 
  \bigl[1+\bar n(\bfs')P\bigl(k,\mu,|\bfs'|\bigr)\bigr]^{-2}},
\label{defPPI}
\end{eqnarray}
where we used the notation $\mu=\hat \bfk\cdot \hat \bfs$, 
$\w(\bfs,\bfk)=\w(\bfs,k,\mu)$ and 
$P(\bfk,|\bfs|)=P(k,\mu,|\bfs|)$. 
In the case $\bar n P(\bfk,|\bfs|)\ll1$, 
we may set $\w(\bfs,\bfk)=1$ and we have
\begin{eqnarray}
   && (2l+1)\langle \Delta {\cal P}_{l}(k)^2 \rangle=
  \langle \Delta {\cal P}_{0}(k)^2 \rangle
  =2{(2\pi)^3\over \Delta V_{k}}
  {1\over
  \int d\bfs' \bar n(\bfs')^2},
\label{defPPI}
\end{eqnarray}
where we assumed the Legendre polynomials are normalized by 
\begin{eqnarray}
  \int_{-1}^1 d\mu {\cal L}_l(\mu) {\cal L}_{l'}(\mu)
  ={2\over 2l+1}\delta_{ll'}.
\end{eqnarray}
Thus the error of the $l$-th moment is smaller than in the 
monopole spectrum by $\sqrt{2l+1}$.

Here we summarize the difference between our method
and that developed by Cole et~al. (1994). Using our
notation, essentially, they first estimate
\begin{eqnarray}
  {\int d\bfs_1 \int d\bfs_2 \w(\bfs_1,\bfk)\w(\bfs_2,\bfk)
  F(\bfs_1)F(\bfs_2)e^{i\bfk\cdot (\bfs_1-\bfs_2)}
  \over
  \int d\bfs \bar n^2(\bfs) \w(\bfs,\bfk)^2}\equiv R(\bfk),
\label{deffff}
\end{eqnarray}
which is same as the right hand side of equation (\ref{defff})
but without ${\cal L}_l(\hat \bfs_h\cdot \hat \bfk)$. 
After averaging $R(\bfk)$ over the angle $\varphi$ (Figure 1), 
and subtracting the shotnoise contribution, 
the anisotropic power spectrum $P(k,\mu)$ can be obtained. 
Then the quadrupole component can be obtained from $P(k,\mu)$
using ${\cal L}_l(\mu)$. In oder to define the 
direction of the line of sight, in their method, 
a Gaussian window function is introduced, and the
above procedure is repeated for the sample with
different windows and the result is averaged.  
Essentially, in their method, the sample is 
divided into subsamples in order to clearly introduce
the quantity $\mu$ associated with a line of sight 
direction.
Thus the method by Cole \etal (1994) first requires computation of 
the anisotropic power spectrum, whilst our method does not. 
The estimation of the multipole spectrum can be performed 
completely parallel with that for the estimation of the monopole 
spectrum. However, in principle, both the methods are equivalent 
in the limit of the distant observer approximation (excepting 
the weight factor).

Next let us consider the estimator for the discrete density field 
of the object catalog. In this case we replace 
${\cal L}_l(\hat \bfs_h\cdot\hat \bfk)$ in equation 
(\ref{defff}) with ${\cal L}_l(\hat \bfs_1\cdot\hat\bfk)$, 
which is valid as long as the distant observer approximation 
is applicable, and then  we may write
\begin{eqnarray}
   R_{l}(\bfk) &=&  A^{-1}\left(
  \sum_{i_1}^N\w(\bfs_{i_1},\bfk)e^{i\bfk\cdot \bfs_{i_1}}
  {\cal L}_l(\hat \bfs_{i_1}\cdot \bfk) 
  -\alpha\sum_{j_1}^{N_s}\w(\bfs_{j_1},\bfk)e^{i\bfk\cdot \bfs_{j_1}}
  {\cal L}_l(\hat \bfs_{j_1}\cdot \bfk)  \right)
\nonumber
\\
  &&\times 
  \left(
  \sum_{i_2}^N\w(\bfs_{i_2},\bfk)e^{i\bfk\cdot \bfs_{i_2}}
  -\alpha\sum_{j_2}^{N_s}\w(\bfs_{j_2},\bfk)e^{i\bfk\cdot \bfs_{j_2}}
  \right),
\end{eqnarray}
where $N$ ($N_s$) is the number of objects of the real (synthetic) catalog, 
respectively, and we have defined $ A=\int d\bfs \bar n(\bfs)^2 \w(\bfs,\bfk)^2$. 
In a similar way, the shotnoise term may be written as
\begin{eqnarray}
   S_{l}(\bfk)=A^{-1}(1+\alpha) \sum_{i}^{N} 
  \w(\bfs_i,\bfk)^2{\cal L}_l(\hat \bfs_{i}\cdot \bfk),
\end{eqnarray}
using the real catalog, or
\begin{eqnarray}
   S_{l}(\bfk)=A^{-1}{(1+\alpha)\alpha} \sum_{j}^{N_s} 
  \w(\bfs_j,\bfk)^2{\cal L}_l(\hat \bfs_{j}\cdot \bfk),
\end{eqnarray}
using the synthetic catalog.
The other possible expression is
\begin{eqnarray}
   S_{l}(\bfk)=A^{-1}\left(\sum_{i}^{N} 
  \w(\bfs_i,\bfk)^2{\cal L}_l(\hat \bfs_{i}\cdot \bfk)
  +\alpha^2\sum_{j}^{N_s} 
  \w(\bfs_j,\bfk)^2{\cal L}_l(\hat \bfs_{j}\cdot \bfk)\right),
\end{eqnarray}
using both the real and synthetic catalogs, as used in 
Cole \etal~(2005).
The choice of the shotnoise estimation does 
not alter our results in the present paper. 

\section{Application to the 2QZ survey}

In this section, we apply the method to the 2dF QSO sample. 
In general, the constraint on the cosmological parameters 
from the QSO sample is not very tight, however, it is useful 
to obtain unique information about the high redshift universe.  
The clustering of the sample has been investigated by the 2QZ
group extensively (e.g., Outram et~al. 2003, Croom et~al.~2005
and references therein). The result from the clustering 
analysis is quite consistent with the prediction of the concordance
$\Lambda$ cold dark matter (CDM) model built up from consideration of other cosmological 
data (e.g., \cite{Spergel},~\cite{Cole},~\cite{Tegmark},~
\cite{Riess},~\cite{Perlmutter}).
The redshift-space distortion in the clustering of the 
2QZ sample has been investigated by Outram et~al. (2004).
In their work, the anisotropic power spectrum 
$P({k_{\scriptscriptstyle \|}},{k_{\scriptscriptstyle \bot}})$
is measured, and the value of $\beta(=\Omega_m^{0.6}/b)$ is
determined, where ${k_{\scriptscriptstyle \|}}$ 
and ${k_{\scriptscriptstyle \bot}}$ are the wave numbers parallel 
and perpendicular to the line of sight direction, respectively.  
In contrast to their approach we focus on the
quadrupole spectrum in the clustering of the 2QZ sample.

The present work is an extension of previous work investigating the 
monopole power spectrum with the 2QZ sample (Yamamoto 2004). Here we briefly 
review the survey details for self-containment. 
The 2QZ survey covers two areas of $5\times75$
deg${}^2$, one in the south Galactic cap (SGC) and the other in the
north Galactic cap (NGC) in the redshift range of less that 3.
The survey area is defined by the equatorial coordinates
from $\alpha=21^{\rm h}40$ to $\alpha=3^{\rm h}15$ and 
$-32.5^{\rm o}\le\delta\le-27.5^{\rm o}$ in the SGC, 
and $9^{\rm h}50\le\alpha\le14^{\rm h}50$ and 
$-2.5^{\rm o}\le\delta\le2.5^{\rm o}$ in the NGC, 
respectively. 

We use the final catalog reported by the 2QZ survey group, 
which is available through the home page http://www.2dfquasar.org/.
Details of the QSO selection of the catalog is also described in 
the reference \cite{Croomtwo}, in which the photometric and 
spectroscopic incompleteness is discussed, as well as the 
angular selection function. The coverage incompleteness is
given by the hole information which is publicly available 
through the home page.  
We generate the random sample through a random process mapping
angular position of a QSO to different position with fixing 
the redshift with an equal probability on each survey area.   
Then we use 10713 and 8442 QSOs in the SGC and the NGC, 
respectively, in the range of redshift $0.2\le z\le 2.2$
incorporating the hole information. 
Outram \etal (2004) used the QSOs in the range of redshift
$0.3\le z\le 2.2$, which is slightly different from our 
choice. This difference does not alter our result because
the number of the QSOs in the range $0.2\le z\le 0.3$ is 
less than  one percent of the total number of the QSOs 
used here. 

Next we explain our theoretical modeling of the power spectrum.
In a redshift survey, the redshift $z$ is the indicator of 
the distance. Therefore we need to assume a distance-redshift
relation $s=|{\bfs}|=s(z)$ to plot a map of objects.
The power spectrum depends on this choice of the radial coordinate 
of the map $s=s(z)$ due to the geometric distortion. 
For the $\Lambda$ CDM model the comoving distance is given by 
\begin{eqnarray}
  r(z,\Omega_m)=\int_0^z{dz'\over H_0\sqrt{\Omega_m(1+z')^3+1-\Omega_m}},
\label{defsz}
\end{eqnarray}
where $H_0=100 h{\rm km/s/Mpc}$ is the Hubble parameter. 
For our fiducial model we adopt the flat $\Lambda$CDM model 
with $\Omega_m=0.3$ Thus our  fiducial model is $s(z)=r(z,0.3)$.

Our theoretical model properly incorporates the light-cone effect 
(redshift-evolution effect of the clustering), the geometric distortion, 
the linear distortion, the Finger of God effects, including the nonlinear 
evolution of the density perturbations.
We model the QSO power spectrum as (e.g., \cite{SMY}, Yamamoto 2002;~2003)
\begin{eqnarray}
  P(\bfk,s(z))=P(k,\mu,s(z))={s(z)^2\over r(z)^2}{ds(z)\over dr(z)}P_{QSO}\left(
{q_{\scriptscriptstyle \|}}\rightarrow k\mu{ds(z)\over dr(z)},
{q_{\scriptscriptstyle \bot}}\rightarrow k\sqrt{1-\mu^2}{s(z)\over r(z)}
\right)
\end{eqnarray}
with 
\begin{eqnarray}
P_{QSO}({q_{\scriptscriptstyle \|}},{q_{\scriptscriptstyle \bot}})
  = b(z)^2\left(1+{d\ln D_1(z)/d\ln a(z)\over b(z)}{{q_{\scriptscriptstyle \|}}^2
  \over q^2}\right)^2 P_{\rm mass}^{\rm Nonlinear}(q,z)
  D[{q_{\scriptscriptstyle \|}} \sigma_P(z)]  
  {\cal D}(\delta z)
\label{nonlin}
\end{eqnarray}
for the nonlinear modeling, 
and 
\begin{eqnarray}
P_{QSO}({q_{\scriptscriptstyle \|}},{q_{\scriptscriptstyle \bot}})
  = b(z)^2\left(1+{d\ln D_1(z)/d\ln a(z)\over b(z)}{{q_{\scriptscriptstyle \|}}^2
  \over q^2}\right)^2 P_{\rm mass}^{\rm Linear}(q,z) 
  {\cal D}(\delta z)
\label{lin}
\end{eqnarray}
for the linear modeling, 
where $q^2={q_{\scriptscriptstyle \|}}^2+{q_{\scriptscriptstyle \bot}}^2$,
$D_1(z)$ is the linear growth factor normalized to $D_1(z=0)=1$, 
$P_{\rm mass}^{\rm Linear}(q,z)$ ($~P_{\rm mass}^{\rm Nonlinear}(q,z)$) is the 
linear (nonlinear) mass power spectrum, $D[{q_{\scriptscriptstyle \|}} \sigma_P(z)]$
is the damping factor due to the Finger of God effect, 
and ${\cal D}(\delta z)$  represents the damping factor due to 
the error in redshift measurement (see equation (\ref{calddmp})).
We model the exponential distribution function for the pairwise 
peculiar velocity, using an approximate formula for the 
mean square velocity dispersion at the large separation
determined through the cosmic energy equation (\cite{Mo}, \cite{magira}, 
\cite{SMJMY}).
In particular for the QSO sample, the error in measuring the
redshift is not negligible, which causes an apparent velocity
dispersion and additional redshift-space distortion. 
We incorporate this effect in modeling the power spectrum 
by multiplying the spectrum with the damping factor 
\begin{eqnarray}
  {\cal D}(\delta z)= \exp\left[-k_{||}^2\left(
  {ds\over dz}\right)^2 < \delta z^2>\right],
\label{calddmp}
\end{eqnarray}
where $k_{\scriptscriptstyle \|}$ 
is the comoving wave number parallel to the line 
of sight direction, $\delta z$ is the variance in the redshift measurement 
error for which we adopt $\delta z=0.0014(1+z)$ (Croom \etal~2005). 

In the modeling of the bias, we consider the scale independent bias 
model of Fry (1996)
\begin{eqnarray}
  b(z)=1+{b_0-1\over D_1(z)},
\label{bias}
\end{eqnarray}
where $b_0$ is the constant bias parameter.

Figure 2 plots our theoretical curves for ${\cal P}_0(k)$ (left panel) and 
${\cal P}_2(k)/{\cal P}_0(k)$ (right panel), to show which effect is important
in the power spectrum. The upper (lower) panel assumes $b_0=1~(b_0=2)$
in the bias model (\ref{bias}). Here we adopt the other 
cosmological parameters $\Omega_m=0.28$, $\Omega_b=0.045$, 
$h=0.7$, $\sigma_8=0.9$ and $n=1$ motivated the WMAP result
(\cite{Spergel}).
In this figure the dashed curve is the linear modeling (\ref{lin}), 
while the solid curve is the nonlinear modeling (\ref{nonlin}). 
Concerning the monopole spectrum ${\cal P}_0$, the linear model 
is very similar to the nonlinear model.
This occurs by a cancellation effect of the two nonlinear 
effects; One is the increase of the amplitude of the real-space
power spectrum at large $k$. The other is the decrease of the 
amplitude due to the damping factor of the Finger of God 
effect $D[{q_{\scriptscriptstyle \|}} \sigma_P(z)]$ and 
the error in redshift measurement ${\cal D}(\delta z)$. 
Concerning the quadrupole divided by the monopole, ${\cal P}_2/{\cal P}_0$, 
the behavior at small $k$ comes from the linear distortion effect,
while the behavior at large $k$ comes from the damping factor 
${\cal D}(\delta z)$  (dashed curve).
In the absence of ${\cal D}(\delta z)$, 
${\cal P}_2/{\cal P}_0$ is constant for $k$ in the linear 
modeling. In the nonlinear model of ${\cal P}_2/{\cal P}_0$  (solid curve), 
the further decrease appears due to the Finger of God effect, thought
the effect is not very significant.
This suggests that the error in redshift measurement is the significant 
effect in the present sample for ${\cal P}_2$. The effect of 
the geometric distortion is negligibly small in our model because 
the fiducial model to plot 
map and the theoretical model is very close. 
Namely, $s(z)$ and $r(z)$ are not very different.

Then we followed the prescription explained in the previous
section for the multipole moments of the power spectrum.
We generated the random catalog with the parameter choice 
$\alpha=1/15$. 
Figure 3 plots the monopole power spectrum (open squares
on the upper panels) and the quadrupole spectrum divided 
by the monopole spectrum (open squares on the lower panels). 
The left and right panels show the results with the SGC and 
NGC samples, respectively. The error bars are estimated from
the expression (\ref{defPPI}). The solid curve in Figure 3
is the theoretical curve of the $\Lambda$CDM model with 
the cosmological parameters same as those of Figure 2, and 
$b_0=1.5$ the best fit value of the 
bias parameter. For small wave numbers
less than $k<0.03 h{\rm Mpc}^{-1}$, the measured quadrupole 
has negative sign. 

Here let us briefly compare our approach with that of 
Outram \etal~(2004). 
As mentioned before, they measured the anisotropic
power spectrum $P(k_{\scriptscriptstyle \|},k_{\scriptscriptstyle \bot})$, 
while we have measured the quadrupole moment ${\cal P}_2(k)$. 
In their analysis, in order to estimate 
$P(k_{\scriptscriptstyle \|},k_{\scriptscriptstyle \bot})$, 
each $5\times 75$ deg${}^2$ strip was divided into $8$ 
regions. And for each of these small region, ${k_{\scriptscriptstyle \|}}$ 
and ${k_{\scriptscriptstyle \bot}}$ were defined by applying the 
distant observer approximation. 
Thus $P({k_{\scriptscriptstyle \|}},{k_{\scriptscriptstyle \bot}})$
were estimated separately in 16 individual $5\times 10$ deg${}^2$
regions. We have also estimated the quadrupole spectrum in these
divided samples, but the method clearly failed to detect
the quadrupole spectrum. As the area of the individual region 
becomes small, the estimated values of the quadrupole spectrum 
become negative. 
This suggests that the quadrupole spectrum with our method is 
sensitive to the shape and the boundary of the survey area 
of the sample. 

Earlier work investigated the constraints from 2QZ on the cosmological
parameters $\Omega_m$, $\Omega_b$ and the equation of state of 
the dark energy $w$,  by fitting to the monopole power spectrum 
(Yamamoto 2002, 2004). In the present work, by adopting the above 
theoretical model, we compare the measured quadrupole spectrum with 
the theoretical prediction. 
Using the bias model (\ref{bias}), for this comparison we define
\begin{eqnarray}
  \chi^2=\sum_{i} {[{\cal P}_l(k_i)^{\rm th}-{\cal P}_l(k_i)^{\rm ob}]^2
  \over \Delta {\cal P}_l(k_i)^2},
\end{eqnarray}
where ${\cal P}_l(k_i)^{\rm ob}$ is the value of the observed 
power spectrum at $k_i$, $\Delta {\cal P}_l(k_i)$ is the variance 
of errors in Figure 3, and ${\cal P}_l(k_i)^{\rm th}$ is the 
theoretical spectrum predictions. Figure 4 shows $\chi^2$ as a function 
of the bias parameter $b_0$. The left and right panels 
show the results for the SGC and NGC samples, 
respectively. Here we fixed the other cosmological 
parameters to be the same as those in Figure 2. 
The degree of freedom is $12$.
In each panel, the dashed curve is the result from 
${\cal P}_0$, the best fit value of $b_0$ is 1.4 (1.5) 
for the SGC (NGC) sample. On the other hand, the
solid curve is the result from the quadrupole, ${\cal P}_2$. 
Because the error of ${\cal P}_2$ is larger than from ${\cal P}_0$, 
the constraint on $b_0$ from ${\cal P}_2$ is weak. However,
we see the minimum of $\chi^2$ of the solid curve 
is located around the minimum of the dashed curve
in the left panel. These are consistent with the
best fit value of $b_0=1.64$ found by the 2QZ group 
though the assumed cosmological model is slightly 
different from ours (Croom \etal~2005).

For the right panel (NGC), the agreement
is not good compared with the SGC sample, but we see 
roughly the same behavior, although there are some troubling 
discrepancies
that deserve further study. We may conclude that the
quadrupole signal is detected in our analysis, 
though the error is rather large. As the reason of the poor 
detection, we infer a contamination of the systematic 
error due to the survey geometry, i.e., the shape and 
the boundary effect of the inhomogeneous survey region.  

Here we discuss the validity of bias model. 
As is mentioned by Croom \etal (2001;2005), the redshift-evolution 
of bias (\ref{bias}) does not fit observational data very much. 
They have given the best fit formula (Croom \etal 2005). 
\begin{eqnarray}
  b(z)=0.53+0.289(1+z)^2.
\label{biasbest}
\end{eqnarray}
The dashed curve in Figure 3 plots the theoretical one 
using this fit formula of the bias evolution. We see
the good agreement between the dashed curve and the
solid curve. Note that the solid curve is the
theoretical curve using the bias model (\ref{bias}).
This agreement comes from the fact that our theoretical
spectrum depends only on a mean value of the bias. 

Finally in this section, we briefly mention the validity
of the distant observer approximation. Cole \etal (1994)
discussed various effects which affect the proper
$\beta$-factor estimation from the redshift-space distortion.
A possible relevant effect here is the large opening
angle on the survey. Cole \etal (1994) found that 
a correction might be needed to be applied for angles greater than 
$30$ degree. However, the range of the redshift of
the 2QZ sample is large. At a mean redshift $z=1.2$,
the wave number $k=2\pi/\lambda$ is less than $0.005~h{\rm Mpc}^{-1}$ 
for $\lambda=s(z)\sin(\theta)$ with 
$\theta=30$ degree. Thus the distant observer approximation
is good in our situation. On the other hand 
for $\theta=5$ degree, $k\simeq0.03~h{\rm Mpc}^{-1}$, 
therefore deviations in the observed quadrupole spectrum away 
from the theoretical prediction may appear around these wavenumbers.

\section{Conclusions}
In summary we have investigated the quadrupole of the 
redshift-space power spectrum, $P_2(k)$, of the 2dF QSO survey. 
First, we have developed an algorithm which estimates the 
quadrupole and monopole spectra in a unified manner. 
The error in a
measurement of the quadrupole spectrum is given analytically.  
We have detected the signal of the redshift distortions in the 
quadrupole spectrum, though the errors are still substantial.

When we used the sample divided into several subsamples, 
the agreement of the estimated spectrum with the theoretical 
model became worse. However, the resulting power 
spectrum is consistent with the theoretical prediction and with previous 
analysis of the 2QZ survey. 
Concerning the bias, however, as discussed by Croom et~al.~(2001;2005), 
the Fry bias model is not the best fitting model 
for the evolution of the QSO bias. To check a more 
realistic model of the bias, we used the formula
(\ref{biasbest}) in the theoretical modeling,
which best fits the data from Croom \etal (2005).
We found that this theoretical curve 
is almost the same as that of 
the Fry's bias model. This is because our theoretical
spectrum depends only on the mean value of the bias.

Constraining the nature of dark energy is one of the 
ultimate purposes of such a clustering analysis. However, 
as demonstrated in (Yamamoto 2004), the 2dF QSO sample 
cannot provide a stringent constraint on the equation of 
state of the dark energy. 
The reason for this is the large shotnoise in the 
QSO distribution since the mean number density is too small. 
The application of the algorithm tested here 
to other, larger, galaxy samples such as SDSS will be 
very interesting (see e.g., Eisenstein \etal~2005, Cole \etal~2005, 
Yahata \etal~2005) and is left to future work.

\vspace{2mm}
{\it Acknowledgments}~~
We thank the anonymous referee for useful comments,
which helped improve the manuscript. 
This work is supported by Grant-in-Aid for Scientific 
Research of Japanese Ministry of Education, Culture, 
Sports, Science, and Technology 15740155.


\newpage
\begin{figure}[t]
\begin{center}
\rotatebox{0}{
\includegraphics[width=10.cm]{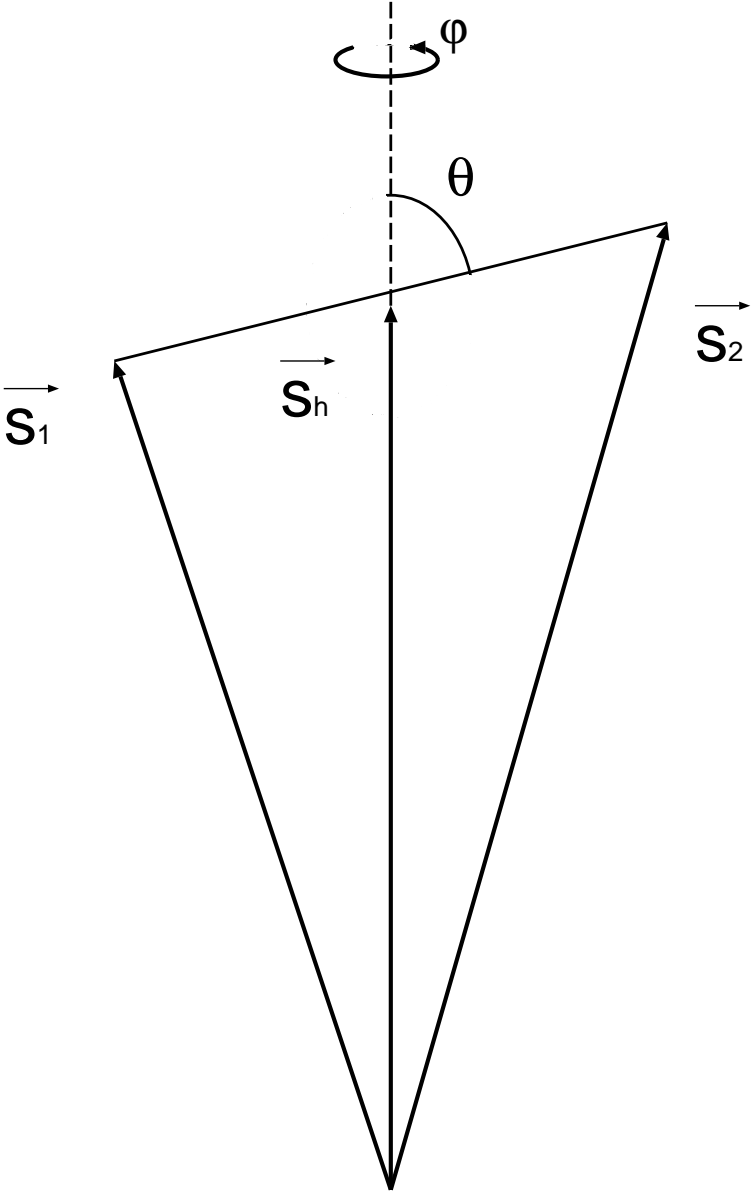}}
\end{center}
\caption{A Sketch for the definition of the variables.
}
\label{sisenfig}
\end{figure}
\newpage
\begin{figure}[t]
\begin{center}
\rotatebox{0}{
\includegraphics[width=14.cm]{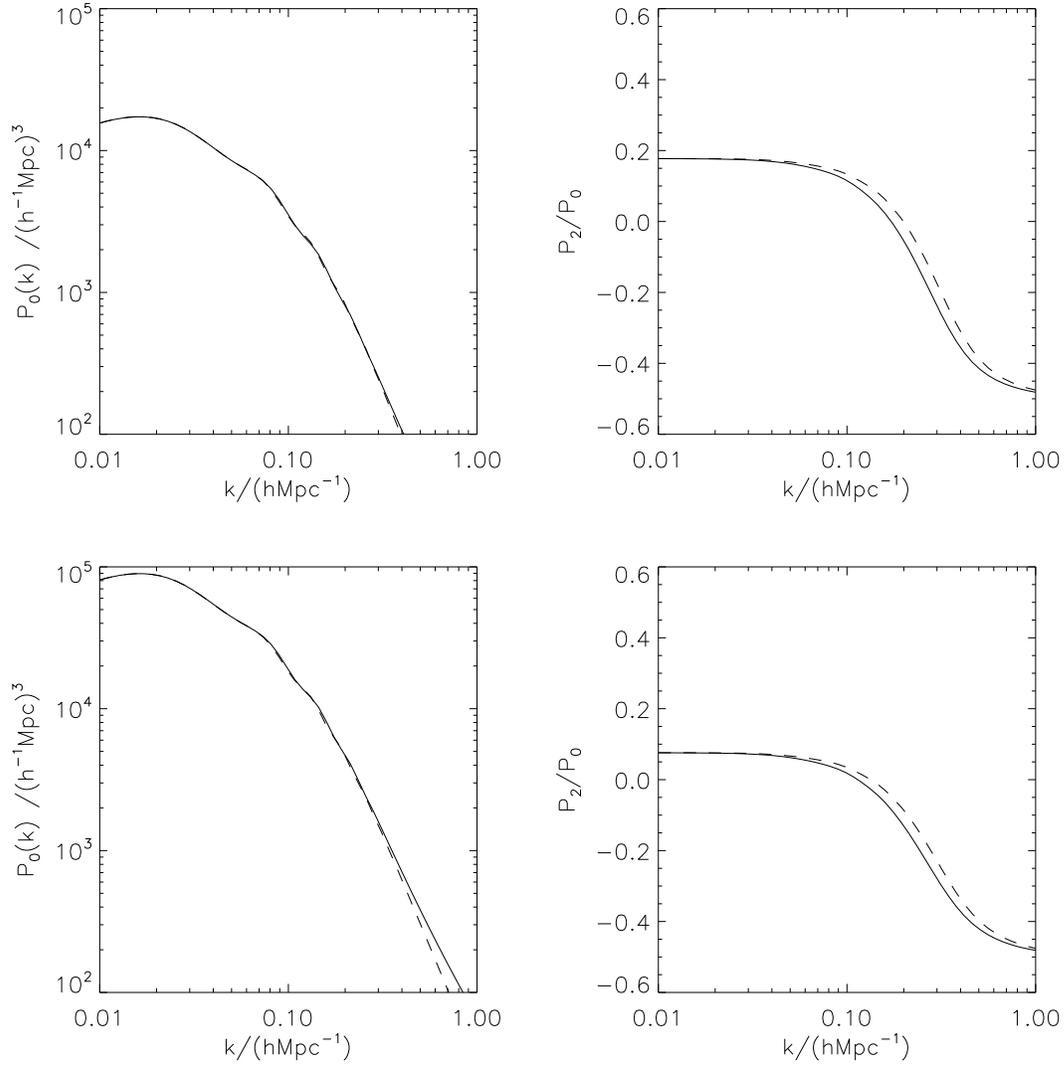}}
\end{center}
\caption{Theoretical prediction for 
the monopole power spectrum ${\cal P}_0(k)$ (left panel) and 
quadrupole ${\cal P}_2(k)/{\cal P}_0(k)$ (right panel). 
The dashed curve is the linear modeling (\ref{lin}), 
while the solid curve is the nonlinear modeling (\ref{nonlin}).
The upper (lower) panels assume $b_0=1~(b_0=2)$
in the bias formula (\ref{bias}).
Theoretical model is the $\Lambda$CDM model with 
the cosmological parameters $\Omega_m=0.28$, $\Omega_b=0.045$, 
$h=0.7$, $\sigma_8=0.9$ and $n=1$. }
\label{figtheory}
\end{figure}

\begin{figure}[t]
\begin{center}
\rotatebox{0}{
\includegraphics[width=14.cm]{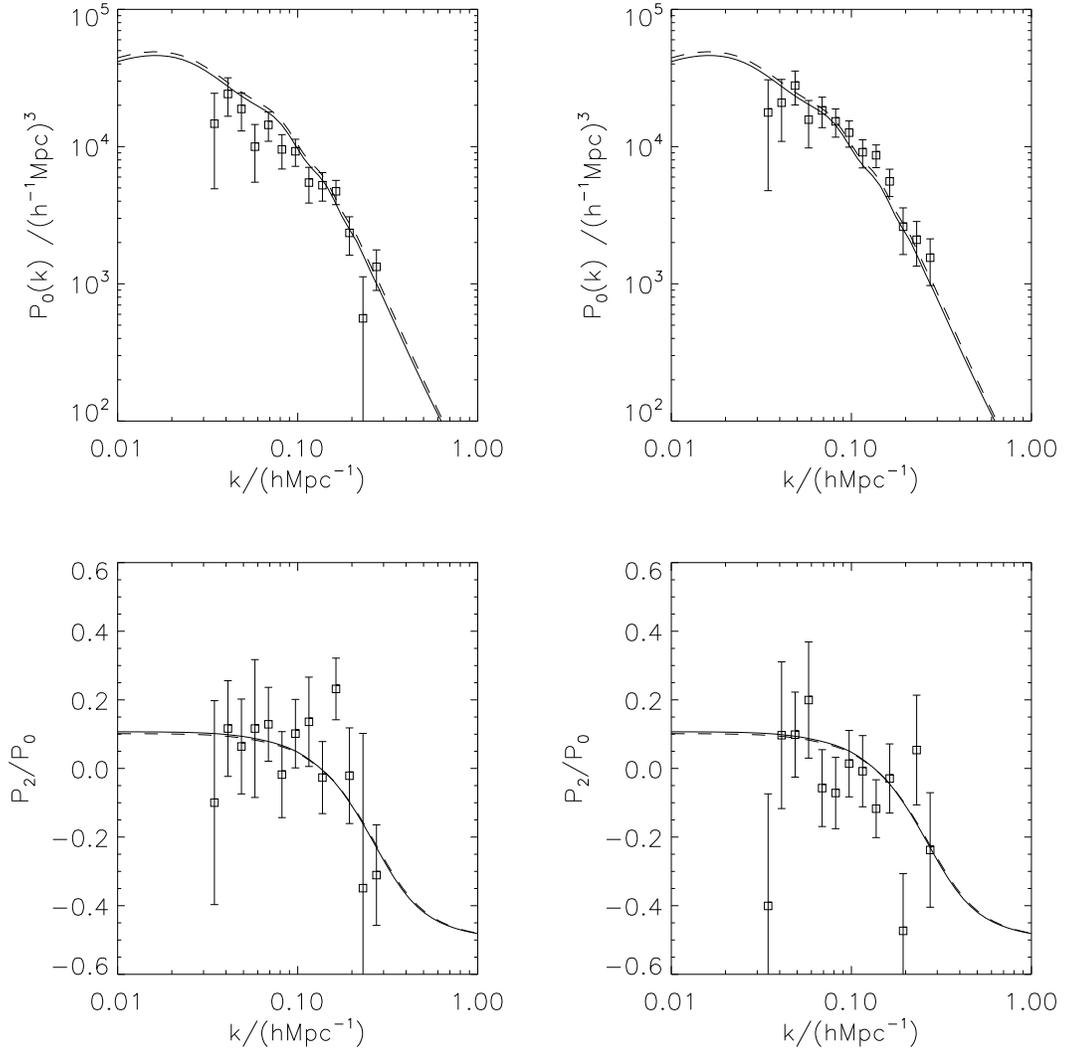}}
\end{center}
\caption{The monopole power spectrum ${\cal P}_0(k)$ (upper panels) and 
quadrupole divided by monopole ${\cal P}_2(k)/{\cal P}_0(k)$ (lower panels) 
from the 2dF QSO sample. 
The left (right) panels are from the SGC (NGC) sample,
respectively. The curves assume the $\Lambda$CDM model with 
the cosmological parameters $\Omega_m=0.28$, $\Omega_b=0.045$, 
$h=0.7$, $\sigma_8=0.9$ and $n=1$.
Concerning the bias model,  the solid curve assumes
(\ref{bias}) with $b_0=1.5$, but the dashed
curve uses (\ref{biasbest}).
}
\label{asca-xte}
\end{figure}

\begin{figure}[t]
\begin{center}
\rotatebox{0}{
\includegraphics[width=14.cm]{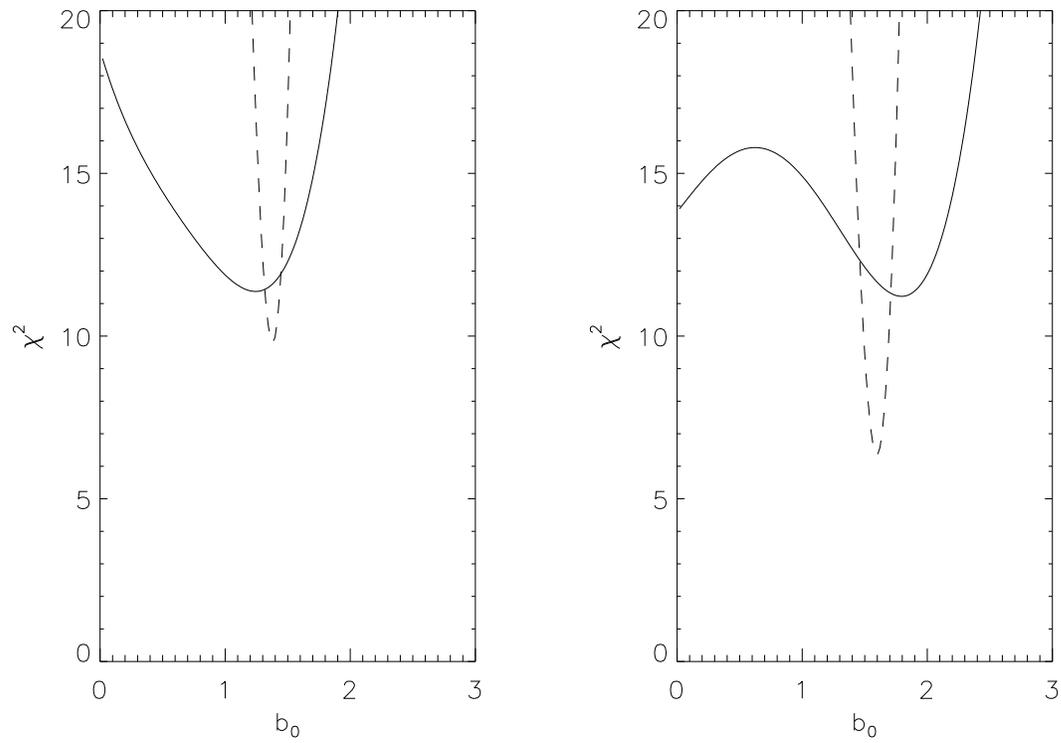}}
\end{center}
\caption{The $\chi^2$ for ${\cal P}_0(k)$ (dashed curve) 
and ${\cal P}_2(k)$ (solid curve) as a
function of the bias parameter $b_0$. 
The left (right) panels are for the South Galactic Cap (North Galactic Cap) sample,
respectively.}
\label{xmm-hardtail}
\end{figure}

\end{document}